# Near infrared light amplification in Gold diffused Silicon-on-Insulator waveguides

S. Stepanov and S. Ruschin
*Department of Physical Electronics, Faculty of Engineering, Tel-Aviv University, Tel-Aviv 69978 ISRAEL*
*E-Mail: ruschin@eng.tau.ac.il*

**Abstract:** We report near infrared optical amplification in gold diffused silicon-on-insulator waveguides by visible optical pumping. More then 30dB/cm gain was measured for a light carrier at a wavelength of 1.55 microns.

There is continuously a great deal of interest in the pursuing of semiconductors as basic materials for lasers and optical amplifiers for optical networks. Most of existing semiconductor lasers and amplifiers are based on direct band gap semiconductor materials. Up of today, Silicon is not considered a favorable candidate for these applications, due to its indirect band gap structure. On the other hand, light stimulated emission properties of GaP [1], also an indirect band-gap semiconductor has been reported and light-emitting diodes (LEDs) based on this material are available. Regarding doped Silicon, a massive effort in order to develop lasers and amplifiers is taking place these days worldwide (see Refs. [2] for a review on the subject).

As is well known, Shockley-Read-Hall (SRH) and Auger recombination are dominant recombination mechanisms in Silicon [3]. SRH recombination (recombination via deep energy levels in Silicon forbidden gap due to sample deformations, doping, etc.) can enhance radiative recombination. The energy level of impurities or other faults will determine the wavelength of the luminescence. Therefore controlled insertion of doping type, traps, or recombination centers can allow the luminescence at specific wavelength. Moreover, if carriers lifetime in a lower energy level, where there is recombination of the free electrons and holes, is less then that of an upper energy level (capture center, conduction band, etc.), it would be possible to realize optical amplification or laser action at specific wavelength.

In this work, we report results which are potentially relevant to the use of Silicon as basic material for lasers and optical amplifiers manufacturing. Free carriers can be generated by applying optical pumping, current injection, application of electrical field (impact ionization mechanism), or combinations of optical pumping with electrical field. These effects are in general combined with radiative recombination.

Radiative recombination mechanisms for Silicon are still not fully clarified in some cases. Most of researchers agree that luminescence in Silicon is a result of the presence of impurities or faults, that create energy levels which are situated in the forbidden band. The impurities can be acquired due to deformation of the silicon sample, or doping of the material. For example, Sauer et al [4] observed dislocation- related photoluminescence in Silicon. Recombination centers were obtained by them via temperature deformation of Silicon samples with different types and concentrations of the doping. Measured photoluminescence spectra ranged from 1.1 to 1.7 μm wavelengths. Additional researchers observed luminescence in gold and silver doped Silicon [5,6]. In those cases, luminescence radiation spectra were well matched with familiar data of the energy level positions for these metals in Silicon forbidden band, specifically, at 0.35eV and 0.34eV, for gold and for silver respectively.

Recently we observed stimulated emission in a bulk Silicon slab by optical pumping, for the first time to our best knowledge [7]. Reports exist on gain in Si derivates including Er[+] doped Silicon [8] and structures not based on the original crystalline structure (nano-crystals, nano-layers etc [9]). Additional mechanism exploited was stimulated Raman scattering [10], a mechanism present in many materials. Our main goal in this work is the investigation of the stimulated emission possibilities in Silicon. Stimulated emission is in the essence of laser action or optical amplification of light signals. Our measurements reported here pertain to this last optical effect. In our previous report [7] we have shown optical amplification in the range of 1.55±0.02μm wavelengths, in Phosphorus- doped Silicon by optical pumping at wavelength of 1.06μm.

For testing of the optical amplification we used Phosphorous-doped ($N \approx 10^{13}$) bond and etchback- Silicon-on-Insulator (BESOI) wafers which were manufactured by Shin-Etsu Hadotai Co., Ltd.. Gold was thermally diffused in top layer of the SOI wafer at different temperature regimes and diffusion times at room atmosphere and pressure. Thickness of the sputtered Gold layer was about 1500A. Diffusion times for separated samples ranged from 30 minutes to 7 hours in 30 minutes steps. Diffusion temperature was varied from 550 to $750^0$C in $50^0$C step for different samples. For some samples we applied fast heating and cooling. Ridge large singlemode waveguides were manufactured after Gold diffusion process. Waveguide geometrical parameters were: 5-μm thickness, 10-μm width, 2-cm length and silicon dioxide buffer layer had 0.5-μm thickness. The top SOI layer was dry etched on 1.5-μm by ICP IRE process for fabrication of the large single mode waveguides.

In the experiment we used either of two infrared lasers to provide the optical signal for amplification. The first one was a tunable laser, ranging in wavelength between 1.527 and 1.576μm and second laser operated at a wavelength of 1.32μm. The signal sources were coupled-in into the waveguide by means of a single-mode fiber. Light coming out from the waveguide was directed into a fast IR photoreceiver by means of a microlens. The amplitude changes of the reference signal infrared light were monitored by an oscilloscope.

Second harmonic of a CW Nd: YAG laser ($\lambda$ =532 nm) was used for optical pumping. Pumping light was shined into the waveguide from above, transversely to the signal laser propagation. The diameter of the output pump laser beam was 3-mm, and a cylindrical converging lens with an 18-mm focal length was used for increasing the incident power density of the pumping light at the test waveguide. The dimensions of the pump light spot impinging the top of the waveguide was 3mm by 10 μm. Only a 3-mm portion of the waveguide was thus illuminated. Pump power ranged from 100mW to 2.5W. A mechanical chopper operating at frequencies ranging from 1Hz to 1KHz and 20% duty cycle was inserted for modulation of the pumping light.

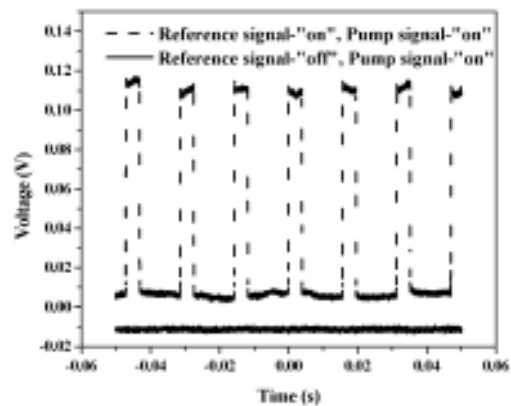

Fig.1 Oscillograms of the infrared signal responsibility in the presence of the pump signal (dotted line) and without infrared signal (straight line)

Figure 1 shows oscillogram of the optical signal response to the pump action. Full line trace at the bottom, corresponds to the situation where the reference signal was switched off while at same time the modulated pump was present. From the bottom trace one concludes that no stray light from the pump was detected by the IR detector at the recorded sensitivity levels. The dashed line shows infrared signal response to the presence of the pump signal. As seen, a 6-7 fold signal amplification was evidenced.

Figures 2 and 3 show transmission of the guided reference infrared signal as a function of pump power, for both, gold diffused and undiffused waveguide samples. We reported the last case in a previous article [11] that dealt with light attenuation by light in SOI- based waveguides. From comparison between two graphs, it can be seen, that indiffused gold changes drastically the optical properties of the Silicon waveguides inducing optical amplification instead of induced absorption of the near infrared light. A maximum gain coefficient of 30dB/cm was obtained at 0.55W pumping power. For pump powers of 1.7W and higher, attenuation was measured instead of amplification, suggesting that free-carrier absorption mechanisms prevailed. From calculations that were presented in our previous work [11], free carrier absorption coefficient can be evaluated to be of about 35dB/cm (data of the Reference [12] taking into account free carrier lifetime for gold-diffused silicon) at a free carrier concentration (about $10^{18}$cm$^{-3}$) induced by 1.7W pump power in similar waveguide without gold. For signal radiation at a wavelength of 1.32-μm, the gain was considerably less (about 6dB/cm).

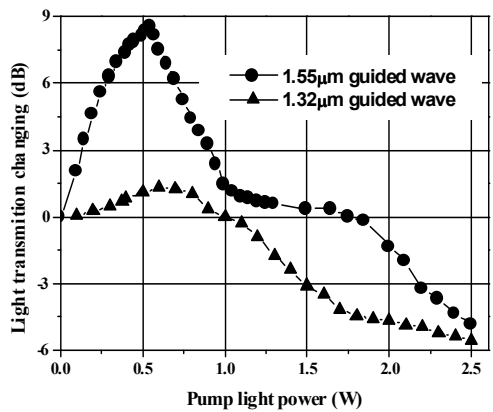 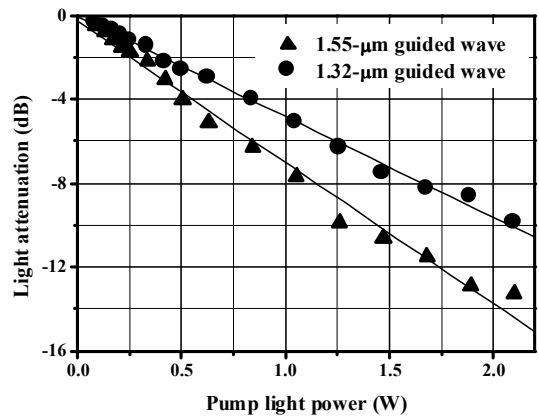

Fig.2. Dependence of the guided reference infrared signal on pump power for Gold diffused Silicon-based waveguide

Fig.3. Dependence of the guided reference infrared signal on pump power for Silicon-based waveguide without Gold-doping

In conclusion, optical gain of more then 30dB/cm at the near infrared was measured in optically pumped gold-diffused Silicon-based waveguides. Possible mechanisms for the amplification effects were discussed in ref. [7]. We hypothesized there that amplification originates from bound states in form of traps, and these mechanisms occur in conjunction with light-induced free-carrier absorption also present in semiconductors. We invest now our effort to the better understanding of both mechanisms, a process that should lead to the development of new type of efficient light sources in silicon.